\documentclass[12pt]{article}
\usepackage{amsmath}
\usepackage{amssymb}
\usepackage{graphicx}
\usepackage{placeins}

\usepackage{titlesec}
\titleformat{\subsection}[runin]{\bfseries}{}{0.0em}{}

\newcommand{\beq}{\begin{equation}}
\newcommand{\eeq}{\end{equation}}
\newcommand{\beqa}{\begin{eqnarray}}
\newcommand{\eeqa}{\end{eqnarray}}

\begin{document}

\def\correspondingauthor{\footnote{Correspondence to \texttt{luca.salasnich@pd.infn.it}.}}

\title{Vortices and antivortices in two-dimensional ultracold Fermi gases}
\author{G. Bighin$^{1}$ and L. Salasnich$^{2,3}\correspondingauthor{}$}

\date{\today}

\maketitle

\begin{center}
$^{1}$IST Austria (Institute of Science and Technology Austria), 
Am Campus 1, 3400 Klosterneuburg, Austria
\\
$^{2}$Dipartimento di Fisica e Astronomia ``Galileo Galilei'' 
and CNISM, Universit\`a di Padova, Via Marzolo 8, 35131 Padova, Italy
\\
$^{3}$Istituto Nazionale di Ottica del Consiglio Nazionale delle Ricerche, 
Via Nello Carrara 1, 50019 Sesto Fiorentino, Italy
\end{center}

\begin{abstract} 
Vortices are commonly observed in the context of classical hydrodynamics: 
from whirlpools after stirring the coffee in a cup to a
violent atmospheric phenomenon such as a tornado,
all classical vortices are characterized by an arbitrary circulation value
of the local velocity field. On the other hand 
the appearance of vortices with quantized circulation represents one of 
the fundamental signatures of macroscopic quantum phenomena. 
In two-dimensional superfluids quantized vortices play a key role in 
determining finite-temperature properties, as the superfluid phase and 
the normal state are separated by a vortex unbinding transition, 
the Berezinskii-Kosterlitz-Thouless transition. 
Very recent experiments with two-dimensional superfluid fermions motivate 
the present work: we present theoretical results based on the renormalization 
group showing that the universal 
jump of the superfluid density and the critical temperature
crucially depend on the interaction strength, providing a strong 
benchmark for forthcoming investigations.
\end{abstract}

Quantized vortices are characterized by a circulation of the velocity 
field quantized in multiples of $\hbar/m^*$, where $\hbar$ is Planck's 
constant and $m^*$ is the mass of a superfluid particle, in the case of a 
bosonic superfluid, or the mass of a Cooper pair, in the case of a fermionic 
superfluid. Quantized vortices are a fundamental feature of superfluid 
and superconducting systems \cite{benfatto9} and have been observed in a wide variety 
of systems, including 
type-II superconductors \cite{tonomura,roditchev,bonevich}, superfluid liquid 
Helium \cite{vinen,bewley}, superfluid liquid Helium nanodroplets 
\cite{gomez1,gomez2}, ultracold gases \cite{hadzibabic,schweikhard}, 
and exciton-polaritons inside semiconductor microcavities 
\cite{lagoudakis,roumpos}.

From a phenomenological standpoint quantized vortices resemble non-quantized 
vortices in classical hydrodynamical systems. The quantization 
of circulation is a peculiar consequence of the existence of an 
underlying \textit{compact} real field, whose spatial gradient 
determines the local superfluid velocity of the system \cite{schakel,wen}. 
This compact real field, the so-called Nambu-Goldstone field, 
is the phase angle of the complex bosonic field which describes, in the case 
of attractive fermions, strongly-correlated Cooper pairs of 
fermions with opposite spins \cite{wen}. 

In two-dimensional (2D) superfluid systems there can not be Bose-Einstein 
condensation and off-diagonal long-range order at finite temperature, 
as a consequence of the Mermin-Wagner-Hohenberg (MWH) theorem 
\cite{mermin,hohenberg,coleman}. Nevertheless a vortex-driven phase 
transition at a finite temperature $T_{\text{BKT}}$ is still present due to the 
Berezinskii-Kosterlitz-Thouless (BKT) mechanism \cite{berezinskii,kosterlitz}. 
Below the critical temperature $T_{\text{BKT}}$ the system is superfluid and 
characterized by bound vortex-antivortex pairs and algebraic long-range 
order. Above $T_{\text{BKT}}$, on the other hand, vortex-antivortex pairs 
unbind, free quantized vortices proliferate, 
and the system loses its superfluid 
properties with exponential decay of coherence. Within this scenario 
it is clear that quantized vortices play a key role in determining the 
finite-temperature properties of a 2D superfluid.

The rapid developments in the realization and manipulation of ultracold 
gases allow for the observation of dilute atomic vapors trapped in 
quasi-two-dimensional configurations. In 2006 the BKT transition 
and the associated unbinding of vortices has been observed in an
atomic Bose gas by Hadzibabic et al.~\cite{hadzibabic}; in this experiment, 
the proliferation of free vortices is directly imaged by letting 
two 2D clouds expand and interfere with each other;
the free vortices can then be counted individually by looking at the number 
of defects in the interference pattern. The same transition was also 
observed by Schweikhard et al.~\cite{schweikhard} in an optical lattice, 
using the usual absorption imaging technique of the vortex cores.
Recent experiments \cite{makhalov,murthy,fenech,boettcher} 
deal with 2D attractive Fermi gases 
in the crossover from the weak-coupling BCS regime of largely overlapping 
Cooper pairs to the strong-coupling BEC regime of composite bosons
and provide motivation for the present theoretical investigation.

\section*{Results} 

\subsection{Single-particle and collective excitations in ultracold Fermi 
superfluids}

In a fermionic superfluid with tunable $s$-wave interaction 
the mean-field theory 
predicts the existence of fermionic single-particle excitations, 
whose low-energy spectrum is
\beq 
E_{sp}(k) = 
\sqrt{\left({\frac{\hbar^2k^2}{2m}} - \mu \right)^2 + \Delta_0^2}  \; , 
\label{disp-sp}
\eeq
where $m$ is the mass of a fermion, 
$\mu$ is the chemical potential of the system, and 
$\Delta_0$ is the pairing energy gap. 
The inclusion of beyond-mean-field 
effects, namely quantum fluctuations of the pairing field, 
gives rise to bosonic collective excitations \cite{diener}, whose low-energy
spectrum across the BCS-BEC crossover is \cite{kurkjian,bighin2}
\beq 
E_{col}(k) = \sqrt{ 2 m c_s^2 \left({\frac{\hbar^2k^2}{2m}}\right) + \lambda 
\left( {\frac{\hbar^2k^2}{2m}} \right)^2 + \gamma 
\left( {\frac{\hbar^2k^2}{2m}} \right)^3 } \; . 
\label{disp-col}
\eeq

These collective excitations are density waves reducing to 
the Bogoliubov-Goldstone-Anderson mode $E_{col}(k)= c_s \hbar k$ 
in the limit of small momenta. 
Here $c_s$ is the speed of sound, while 
$\lambda$ and $\gamma$ are parameters taking into account the increase 
of kinetic energy due to the spatial variation of the density and depend on the 
strength of the attractive interaction: in the deep BEC regime one 
finds $\lambda=1/4$ and $\gamma=0$  such that $E_{col}(k)=\hbar^2k^2/(4m)$ 
for large momenta. It has been demonstrated that the inclusion 
of collective excitations in the equation of state, as briefly outlined in the Methods and derived in Refs. \cite{bighin,salasnich3},
recovers the correct composite boson limit at zero temperature \cite{salasnich3}, also providing qualitatively 
good results for many observable quantities across the whole 
crossover \cite{he,bighin}; we follow this approach in the present work.

The superfluid (number) density $n_s$ of the two-dimensional (2D) fermionic 
system can be written as 
\beq 
n_s = n - n_n = n - n_{n,sp} - n_{n,col} \; , 
\eeq
where $n$ is the 2D total number density and $n_{n}=n_{n,sp}+n_{n,col}$ 
is the 2D normal density due to both single-particle and collective 
elementary excitations \cite{landau}. For a uniform superfluid system
at zero temperature $n_{n}=0$ and $n_s=n$. As the temperature is increased, 
the normal density $n_n$ increases 
monotonically and, correspondingly, the superfluid density $n_s$ decreases. 
According to the Landau's approach 
\cite{landau,fetter}, the two contributions to the normal density read 
\beq
n_{n,sp} = \beta \int \frac{\mathrm{d}^2 \mathbf{k}}{(2 \pi)^2} 
{\frac{\hbar^2k^2}{m}} \frac{e^{\beta E_{sp} (k)}}{(e^{\beta E_{sp} (k)} + 1)^2}
\label{eq:nnf}
\eeq
and
\beq
n_{n,col} = \frac{\beta}{2} \int \frac{\mathrm{d}^2 \mathbf{q}}{(2 \pi)^2} 
{{\hbar^2q^2}{m}} \frac{e^{\beta E_{col} (q)}}{(e^{\beta E_{col} (q)} - 1)^2} \;.
\label{eq:nnb}
\eeq
where $\beta=1/(k_BT)$, $k_B$ the Boltzmann constant and 
$T$ the absolute temperature. The superfluid density $n_s$ can also be 
inferred from the coefficient governing phase fluctuations in an effective 
action for the system \cite{devreese}; it turns out that for a 
Gaussian-level action this approach is equivalent to setting 
$n_s = n - n_{n,sp}$, ignoring the contribution from collective 
excitations to the superfluid density; this contribution, however, 
will turn out to be fundamental in 
the strong coupling regimes that have become recently 
accessible \cite{murthy}. 

More generally, in the extreme BCS (BEC) limit only the fermionic (bosonic) excitations contribute to the total superfluid density. As already discussed in Ref. \cite{bighin}, the present approximation, considering the fermionic and bosonic excitations as separate, neglects the Landau damping that hybridizes the collective modes with the single-particle excitations \cite{griffin}. It should be stressed, however, that the Landau damping is absent at $T=0$, making our approximation reliable in the low-temperature limit. Moreover we also discussed \cite{bighin} that Landau damping would affect the bosonic contribution $n_b$ in the BCS region, where the physics is dominated by the fermionic contribution. This interplay makes the Landau damping less relevant as far as the present work is concerned, justifying the present choice of approximation.

The effective low-energy Hamiltonian of a fermionic superfluid can be 
recast as that of an effective 2D XY model \cite{babaev,loktev,hadzibabic2}:
\beq 
H = \frac{J}{2} \int \mathrm{d}^2{\bf r} \, ( {\boldsymbol \nabla} 
\theta({\bf r}) )^2 \; , 
\label{ham}
\eeq
having introduced the pairing field $\Delta(\mathbf{r}) 
= | \Delta(\mathbf{r}) | e^{\mathrm{i} \theta(\mathbf{r})}$ 
with $\theta({\bf r})$ the so-called Nambu-Goldstone field \cite{wen}.
The phase stiffness $J$ is a function 
of the fermion-fermion attractive strength and of the temperature; 
it measures the energy cost associated to space variation 
in the phase angle $\theta({\bf r})$ of the pairing field. 
Moreover the phase stiffness $J$ is proportional to the 
superfluid number density $n_s$, namely \cite{fisher}
\beq 
J = {\frac{\hbar^2}{4 m}} n_s \; . 
\label{eq:barej}
\eeq
The compactness of the phase angle field $\theta({\bf r})$ 
implies that $\oint_\mathcal{C} {\boldsymbol \nabla} \theta({\bf r}) \cdot 
\mathrm{d}{\bf r} = 2\pi q $ for any closed contour $\mathcal{C}$. 
Here $q=0,\pm 1,\pm 2,...$ is the 
integer number associated to the corresponding 
quantum vortex (positive $q$) or antivortex (negative $q$).
Consequently the circulation of the superfluid velocity
${\bf v}({\bf r})=(\hbar/m^*) \nabla \theta({\bf r})$
is quantized according to
$ \oint_\mathcal{C} {\bf v} \cdot d{\bf r} = (2 \pi \hbar / m^*) \ q $
where $m^*=2m$ is the mass of a Cooper pair.
Formally, one can rewrite the phase angle as follows 
\beq 
\theta({\bf r}) = \theta_0({\bf r}) + \theta_v({\bf r}) \; , 
\eeq
where $\theta_0({\bf r})$ has zero circulation (no vortices) while 
$\theta_v({\bf r})$ encodes the contribution of quantized vortices. 
Consequently, the Hamiltonian in Eq. (\ref{ham}) can be rewritten 
\cite{nagaosa} as $H = H_0 + H_v$ where 
$
H_0 = J/2 \int \mathrm{d}^2{\bf r} \, ( {\boldsymbol \nabla} 
\theta_0({\bf r}) )^2 
$
is the Hamiltonian of density oscillations, while 
\beq 
H_v = \sum_{i\neq j} V({\bf r}_i -{\bf r}_j) q_i q_j  
- \sum_j \mu_c \, q_j^2 \; , 
\eeq
is the Hamiltonian of quantized 
vortices located at position ${\bf r}_i$ with quantum numbers $q_i$, 
interacting through a 2D Coulomb-like potential 
\beq 
V(r) = - 2\pi J \ln{\left({\frac{r}{\xi}}\right)} \; , 
\label{pot2d}
\eeq
where $\xi$ is the healing length, i.e. the cutoff length 
defining the vortex core size, and $\mu_c$ the energy associated 
to the creation of a vortex \cite{nagaosa,benfatto}. 

\subsection{Renormalization group analysis for a Fermi superfluid}

The total number of quantized vortices varies as a function of the temperature: 
at zero temperature there are no vortices, however as the temperature 
increases vortices start to appear in vortex-antivortex pairs. 
Due to the logarithmic energy cost the pairs are bound at low temperature, 
until at the critical temperature $T_{\text{BKT}}$ an unbinding transition 
occurs above which a proliferation of free vortices 
and antivortices is observed \cite{kosterlitz}. Vortex-antivortex 
pairs with small separation distance can screen the potential 
in Eq. (\ref{pot2d}) between a vortex-antivortex pair with larger distance $r$;
as a consequence, the phase stiffness $J$ and the vortex energy $\mu_c$ 
are renormalized \cite{nelson}. In particular analyzing the effect 
of increasing the spatial cutoff $\xi$, thereby excluding 
vortex-antivortex configurations with distance smaller than $\xi$, 
Nelson and Kosterlitz obtained the renormalization 
group equations \cite{nelson,nagaosa,benfatto}
\beq
\begin{cases}
\frac{\mathrm{d}}{\mathrm{d} \ell} K(\ell) = - 
4\pi^3 K(\ell)^2 y(\ell)^2 + O(y^3) \\
\frac{\mathrm{d}}{\mathrm{d} \ell} y(\ell) = 
\left( 2 - \pi K(\ell) \right) y(\ell) + O(y^2)
\end{cases}
\label{piopio}
\eeq
subsequently extended by Amit \cite{amit} and Timm \cite{timm}, 
including next-to-leading order terms, in order to describe higher 
vortex densities
\beq
\begin{cases}
\frac{\mathrm{d}}{\mathrm{d} \ell} K(\ell) = 
- 4\pi^3 K(\ell)^2 y(\ell)^2 \left( 2- \frac{K(\ell)}{K(0)} \right) + O(y^3) \\
\frac{\mathrm{d}}{\mathrm{d} \ell} y(\ell) = 
\left( 2 - \pi K(\ell) \right) y(\ell) - 2 \pi^2 y^3(\ell) + O(y^4)
\end{cases}
\label{piopio2}
\eeq
for the running variables $K(\ell)$ and $y(\ell)$, as a function of the 
adimensional scale $\ell$ subjected to the initial conditions
$K(0) = {\beta J}=\beta \hbar^2 n_s/(4m)$ and 
$y(0) = \exp(-\beta \mu_c)$. As discussed in \cite{benfatto}, 
the choice of $\mu_c$, slightly  affecting the final results, 
is still an open problem. The 2D XY model on a lattice 
with a finite difference approximation of spatial derivatives  
implies $\mu_c=\pi^2 J/2$ \cite{nagaosa}. However, for the 2D XY model 
in the continuum it has been suggested $\mu_c \simeq \pi^2 J/4$ 
within the Ginzburg-Landau theory of superconducting films 
\cite{nylen,khawaja,zhang} and, more recently, 
$\mu_c \simeq 3 J/\pi$ within a phenomenological 
BCS approximation \cite{benfatto}. In our study of the 2D BCS-BEC crossover 
with Eqs. (\ref{piopio}) we adopt $\mu_c = \pi^2 J / 4$, that is currently 
the most rigorous choice for superconductors and 
superfluids \cite{nylen,khawaja,zhang}. 
The renormalized phase rigidity $J^{(R)}$ and the renormalized 
vortex energy \cite{khawaja,nagaosa} $\epsilon_c^{(R)}$ are then 
derived from $K(\infty)$ and $y(\infty)$. 
Finally, one obtains the renormalized superfluid density as 
\beq 
n_{s}^{(R)} = {\frac{4m}{\hbar^2}} {\frac{K(\infty)}{\beta}} \; .
\label{eq:renormalizer_ns}
\eeq
The renormalized superfluid density $n_{s}^{(R)}$ is a monotonically 
decreasing function of the temperature, as is the bare (unrenormalized) 
superfluid density $n_s$; however, while $n_s$ is continuous, $n_s^{(R)}$ 
jumps discontinuously from a finite value to zero as the temperature 
reaches the BKT critical temperature $T_{\text{BKT}}$, implicitly defined 
by the Kosterlitz-Nelson condition \cite{nelson}:
\beq
k_B T_{\text{BKT}} = \frac{\hbar^2 \pi}{8 m} n_s^{(R)} (T_{\text{BKT}}^{-}) \; .
\label{eq:kncondition}
\eeq

Let us verify the validity of the perturbative treatment of the 
renormalization group analysis. Combining Eq. (\ref{eq:barej}), 
Eq. (\ref{eq:kncondition}) and the definition of $\mu_c$ one readily 
sees that the expansion parameter $y$ is a monotonically increasing 
function of the temperature, increasing from $y(\ell=0)=0$ at $T=0$, 
to $y(\ell=0)=\exp(- \pi/2) \simeq 0.208$ at $T=T_{BKT}$. This fact 
suggests that even the leading-order renormalization group in 
Eq. (\ref{piopio}) could give accurate results for the present problem, 
and in fact including the next-order correction as in Eq. 
(\ref{piopio2}) modifies our estimates of the critical 
temperature $T_{BKT}$ by at most $1.5\%$ over the whole crossover (see below), 
confirming the validity of the renormalization group analysis.

\begin{figure}
\centering
\includegraphics[width=0.8\textwidth]{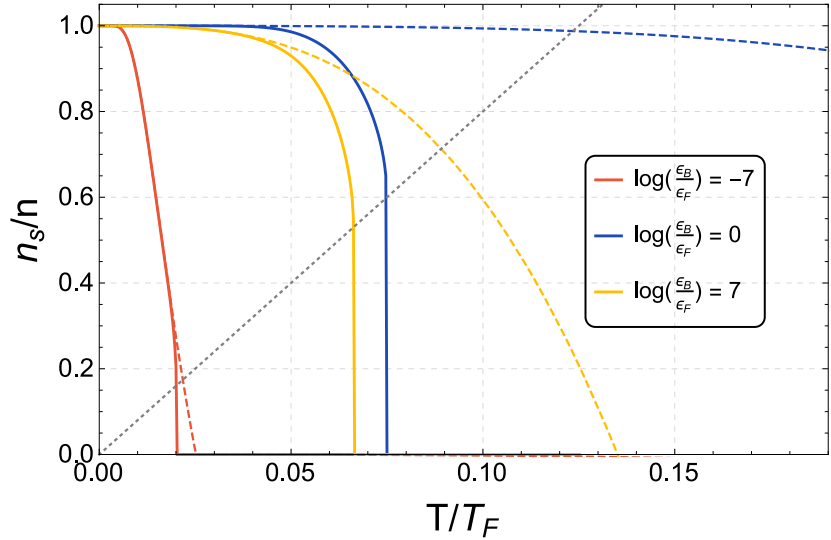}
\caption{The superfluid density, for three different values of the 
interaction, ranging from the BCS to the BEC regime. The solid lines 
represent the results of the renormalization group analysis which is the 
central point of the present paper, whereas the dashed lines represent 
the unrenormalized result obtained from the single-particle and collective 
contributions to superfluid density, as done in Ref. \cite{bighin}. 
The gray dotted line corresponds to the Nelson-Kosterlitz condition 
in Eq. (\ref{eq:kncondition}), showing that the contribution from the 
renormalization group lowers the critical temperature. The universal jump 
as a consequence of the BKT appears for every value of the interaction; 
however the size of the universal jump and the related critical temperature 
are strongly interaction-dependent.}
\label{fig:rhos}
\end{figure}

In Fig. \ref{fig:rhos} we report the renormalized and bare superfluid 
densities for three different values of the interacting strength, in the BCS, 
intermediate and BEC regimes. The renormalization of superfluid density as 
analyzed in Eq. (\ref{eq:renormalizer_ns}) is more evident at higher 
temperatures, as the universal jump defined by Eq. (\ref{eq:kncondition}) 
is approached. We also note that, although always a monotonically decreasing 
function of the temperature, the superfluid density exhibits different 
behaviors across the BCS-BEC crossover, as it can be dominated either by 
fermionic, single-particle excitations, in the weakly-coupled regime, 
or by bosonic, collective excitations, in the strongly-coupled regime.

\subsection{Phase diagram} The finite-temperature phase diagram in the 
present 2D case is profoundly different with respect to a three-dimensional 
Fermi gas as a result of the BKT mechanism just analyzed and also as a result 
of the MWH theorem \cite{mermin,hohenberg,coleman} prohibiting symmetry 
breaking at any finite temperature. These striking qualitative differences 
render a complete analysis of the 2D Fermi gas compelling both from the 
theoretical and experimental point of view.  Let us briefly discuss the 
three possible phases \cite{wen}:
\begin{itemize}
\item \textbf{Condensation}: a 2D superfluid system exhibits 
condensation and off-diagonal long-range order (ODLRO) only strictly 
at $T=0$: this zero-temperature regime is characterized by a 
non-decaying phase-phase correlator $ \langle e^{\mathrm{i} 
\theta(\mathbf{r})} e^{\mathrm{i} \theta(0)} \rangle \sim C$, where $C$ 
is independent of $\mathbf{r}$, and by a finite condensate density 
\cite{salasnich2}.
\item \textbf{Quasi-condensation:} the intermediate phase from 
$T=0^+$ to $T_{\text{BKT}}$ is characterized by the phase-phase correlator 
showing \textit{algebraic} quasi-long-range order 
$ \langle e^{\mathrm{i} \theta(\mathbf{r})} e^{\mathrm{i} \theta(0)} \rangle 
\sim |\mathbf{r}|^{-\alpha} $ for an opportune exponent $\alpha>0$. Although 
the condensate density is strictly zero, a finite superfluid density is 
still present. 
\item \textbf{Normal state}: finally for $T>T_{\text{BKT}}$ the system enters 
the normal phase, characterized by the exponential decay of the phase-phase 
correlator, $ \langle e^{\mathrm{i} \theta(\mathbf{r})} e^{\mathrm{i} \theta(0)} 
\rangle \sim \exp (-|\mathbf{r}|/\xi) $ and by the absence of both 
superfluid and condensate.
\end{itemize}

\begin{figure}
\centering
\includegraphics[width=0.8\textwidth]{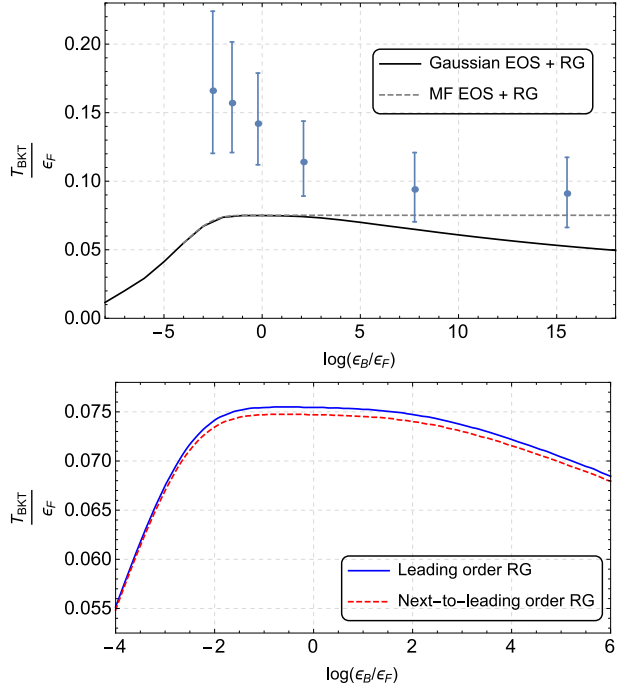}
\caption{The Berezinskii-Kosterlitz-Thouless critical temperature 
as a function of the bound-state binding energy $\epsilon_B$. 
{\bf Upper panel}. The dashed line is the result of renormalization 
group (RG) analysis, i.e. Eqs. (\ref{piopio}), of the mean-field results, 
whereas the solid line uses the Gaussian theory 
as the starting point. The blue dots represent experimental data 
from Ref. \cite{murthy}. The decrease of the critical temperature in 
the BCS and BEC limits is due to single-particle excitations and 
collective excitations contributing to superfluid density, respectively. 
This interplay results in a higher BKT critical temperature in the 
intermediate regime, i.e. when $\epsilon_B \sim \epsilon_F$. 
It is important to note that experimental data may be affected by 
systematic errors, as analyzed in the main text. 
{\bf Lower panel}. Comparison between the Kosterlitz-Thouless 
renormalization group (RG) equations (\ref{piopio}) and the next-to-leading order 
RG equations (\ref{piopio2}). Here, in both cases the bare superfluid density 
is calculated within the Gaussian theory.}
\label{fig:tbkt}
\end{figure}

The gray dashed line in Fig. \ref{fig:rhos} corresponds to the 
Kosterlitz-Nelson condition in Eq. (\ref{eq:kncondition}), identifying 
the critical temperature $T_{\text{BKT}}$, separating the normal state from 
the phase characterized by quasi-condensation. A determination of the 
critical temperature across the whole crossover is reported in the upper 
panel of Fig. \ref{fig:tbkt}, black solid line. 
The rapid decrease of $T_{\text{BKT}}$ 
approaching both the BCS and the BEC limit is a consequence of the fermionic 
single-particle excitations and bosonic collective excitations dominating 
the superfluid density, respectively, rapidly decreasing the normal density 
as either limit is approached. A consequence of this interplay is that the 
critical temperature is higher in the intermediate regime 
($\epsilon_B \sim \epsilon_F$), where the superfluid density is neither 
fermion-dominated nor boson-dominated.

The current approach, involving the inclusion of Gaussian fluctuations 
in the equation of state, the inclusion of bosonic collective excitations 
in the superfluid density along with a renormalization group 
analysis is able to reproduce the downward trend as the interaction 
get stronger; the renormalization group analysis on top of a mean-field 
theory would not have been sufficient to reproduce the correct trend, 
as shown by the gray dashed line in the upper panel of 
Fig. \ref{fig:tbkt}. In other words, 
as also observed elsewhere \cite{salasnich3,he,bighin}, Gaussian 
fluctuations are required in order to correctly describe the physics of 
an interacting Fermi gas in the strongly-coupled limit.

The underestimation of experimental data \cite{murthy}, as observed 
in Fig. \ref{fig:tbkt} may have different causes:

\begin{itemize}
\item 
In the experiment there is a harmonic trap also in the planar direction. 
The effect of the trap can enhance the critical temperature with respect 
to the uniform system, as found in the 3D case by 
Perali {\it et al.} \cite{peraliet,marsiglio}.  
\item 
It has been argued \cite{matsumoto} that the algebraic decay of the 
first-order correlation function, presented in Ref. \cite{murthy} as 
the signature of the superfluid state, could be interpreted in terms 
of the strong-coupling properties of a normal-state. Experimental 
data in Ref. \cite{murthy} would then overestimate $T_\text{BKT}$.
\item 
The determination of the critical temperature may be affected by 
three-dimensional effects, the superfluid not being trapped in a 
strictly 2D configuration.
\item 
On more general grounds one may argue that $T_\text{BKT} > 0.125 \epsilon_F$, 
as experimentally observed in the BCS regime, is not compatible with 
the Kosterlitz-Nelson condition, signaling different mechanisms 
at work \cite{bighin}.
\end{itemize}

For the sake of completeness, in the lower panel of Fig. \ref{fig:tbkt} 
we plot the BKT critical temperature $T_{BKT}$ obtained with 
the Kosterlitz-Thouless renormalization group equations (\ref{piopio}) 
and the generalized renormalization group equations (\ref{piopio2}), 
starting with the bare superfluid density 
derived from the Gaussian theory. As previously stressed the 
relative difference in the determination of $T_{BKT}$ 
is below $1.5\%$ in the whole crossover. Moreover, the figure shows that 
this very small difference is larger in the intermediate coupling regime ($\epsilon_B \sim \epsilon_F$). 

\section*{Discussion}

In the present work we have analyzed the role of vortex proliferation 
in determining the finite-temperature properties of a 2D interacting Fermi 
gas, throughout the BCS-BEC crossover, as the fermion-fermion interaction 
strength is varied. Using the Kosterlitz renormalization group equations 
we have shown that the bare superfluid density is renormalized as 
the vortex-vortex potential is screened at large distances. 
The renormalization of superfluid density lowers the BKT critical 
temperature, correctly reproducing the trend observed in experimental 
data through a non-trivial interplay between the single-particle and collective excitations.
As previously pointed out, and analyzed in Ref. 
\cite{matsumoto}, currently available experimental data may overestimate 
the BKT critical temperature of the uniform system 
and our theoretical predictions are providing a benchmark for forthcoming experiments.

\section*{Methods}

\subsection{Equation of state} The pairing gap $\Delta_0$ and the chemical 
potential $\mu$ are calculated self-consistently by jointly solving the gap 
and number equation, as done e.g. in Refs. \cite{he,bighin}. The Gaussian 
pair fluctuations scheme \cite{hu,klimin} has been adopted which, as opposed 
as the Nozi\`eres-Schmitt-Rink \cite{nozieres} approach, leads to finite, 
converging results in 2D. The spectrum of fermionic and collective 
excitations, $E_{sp}(k)$ and $E_{col}(q)$ as introduced in Eqs. (\ref{disp-sp}) and (\ref{disp-col}), are calculated 
by looking at the poles of the respective Green's functions, as analyzed 
e.g. in Ref. \cite{diener}. Accordingly, the corresponding thermodynamical
grand potential has two contributions, namely the mean-field, fermionic part
\beq
\Omega_F = \frac{2}{\beta} \sum_\mathbf{k} \ln (1+e^{-\beta E_{sp} (k)})
\eeq
and the bosonic part
\beq
\Omega_B = \frac{1}{\beta} \sum_\mathbf{q} \ln (1-e^{-\beta E_{col} (q)}) \; .
\eeq
We stress that $\Omega_F$ accounts for the mean-field description of a tunable
Fermi gas, whereas $\Omega_B$ includes the  contribution of density waves on top of the
mean-field picture.

\subsection{Data availability}

The data is available upon request. Requests should be addressed 
to either author.

\section*{Acknowledgments}
The authors thank L. Benfatto, P. A. Marchetti, C. S\`a de Melo, I. Nandori
G. Strinati, F. Toigo, and A. Trombettoni for fruitful discussions.

\section*{Author contributions} 
G.B. and L.S. jointly defined the project and derived analytical formulas. 
G.B. carried out the numerical calculations. 
G.B. and L.S. analyzed the data and wrote the paper.

\section*{Additional information} 
\subsection{Competing financial interests}
Authors have no competing financial interests.

\end{document}